\documentclass[]{interact}

\usepackage{placeins}
\usepackage{epstopdf}
\usepackage{comment}
\usepackage{units}
\usepackage[caption=false]{subfig}
\usepackage{xcolor}
\usepackage[numbers,sort&compress]{natbib}
\bibpunct[, ]{[}{]}{,}{n}{,}{,}
\renewcommand\bibfont{\fontsize{10}{12}\selectfont}
\makeatletter
\def\NAT@def@citea{\def\@citea{\NAT@separator}}
\makeatother

\theoremstyle{plain}

\theoremstyle{definition}

\theoremstyle{remark}

\begin{document}


\title{Fast biological imaging with quantum-enhanced Raman microscopy}

\author{ 
\name{Alex Terrasson\textsuperscript{a}, Nicolas P. Mauranyapin\textsuperscript{a}, Catxere A. Casacio\textsuperscript{b}, Joel Q. Grim \textsuperscript{c}, Kai Barnscheidt\textsuperscript{d}, Boris Hage \textsuperscript{d}, Michael A. Taylor \textsuperscript{e}, W.~P. Bowen\textsuperscript{a}\thanks{CONTACT W.~P. Bowen. Email: w.bowen@uq.edu.au}}
\affil{\textsuperscript{a} Australian Research Council Centre of Excellence in Quantum Biotechnology, School of Mathematics and Physics, University of Queensland, Australia. \textsuperscript{b} University of Surrey, UK. \textsuperscript{c} Naval Research Laboratory, USA. \textsuperscript{d} Institut für Physik, Universität Rostock, Rostock, Germany. \textsuperscript{e} Department of Anatomy, University of Otago, Dunedin, New Zealand}
}

\maketitle

\begin{abstract}

Stimulated Raman scattering (SRS) microscopy is a powerful label-free imaging technique that probes the vibrational response of chemicals with high specificity and sensitivity. High-power, quantum-enhanced SRS microscopes have been recently demonstrated and applied to polymers and biological samples. 
Quantum correlations, in the form of squeezed light, enable the microscopes to operate below the shot noise limit, enhancing their performance without increasing the illumination intensity. This addresses the signal-to-noise ratio (SNR) and speed constraints introduced by photodamage in shot noise-limited microscopes. 
Previous microscopes have either used single-beam squeezing, but with insufficient brightness to reach the optimal ratio of pump-to-Stokes intensity for maximum SNR, or have used twin-beam squeezing and suffered a 3 dB noise penalty. 
Here we report a quantum-enhanced Raman microscope that uses a bright squeezed single-beam, enabling operation at the optimal efficiency of the SRS process. The increase in brightness leads to multimode effects that degrade the squeezing level, which we partially overcome using spatial filtering.

We apply our quantum-enhanced SRS microscope to biological samples, and demonstrate quantum-enhanced multispectral imaging of living cells. The imaging speed of 100$\times$100 pixels in 18 seconds allows the dynamics of cell organelles to be resolved. 
The SNR achieved is compatible with video rate imaging, with the quantum correlations yielding a 20\% improvement in imaging speed compared to shot noise limited operation.

\end{abstract}


\section{Introduction}

Over the past decades, precision microscopy has advanced dramatically, including the capability to image biological samples with near-atomic resolution \cite{sahl2017fluorescence,sigal2018visualizing}, conduct imaging deep into scattering tissue \cite{helmchen2005deep,huisken2004optical} and perform three-dimensional imaging of the dynamics of cells \cite{valm2017applying,lu2019lightsheet}. 
Following these advances, the performance of several types of microscopes is now limited because the intense optical filed required for increased performance may cause photodamage. The constraint this places on illumination intensity limits the strength of signals that can be extracted from the sample.

The SNR can still be improved by reducing the noise floor. However, without quantum correlations, the noise floor is fundamentally constrained by the shot noise limit, which many microscopes have already reached \cite{bowen2023quantum}.
Together, photodamage and shot noise floor therefore limit the speed, contrast, and sensitivity of conventional microscopes.


Among the microscopy techniques limited by photodamage, we focus on stimulated Raman scattering (SRS) microscopy, a label-free, highly sensitive technique that relies on the Raman effect to fingerprint the chemical composition of samples
\cite{cheng2015vibrational,camp2015chemically}. In the Raman effect, an incident pump photon scatters inelastically from a molecule, resulting in the emission of a lower energy Stokes photon after exciting a chemical bond vibration. The frequency difference, known as the \textit{Raman shift}, matches the vibrational frequency of the chemical bond, allowing to recognise a molecule. SRS enhances the weak Raman scattering process by stimulating the energy transfer with resonant driving at both the Stokes and pump frequencies. The energy transfer from the pump to the Stokes field is proportional to the chemical bond concentration, allowing quantitative and label-free chemical fingerprinting of the sample. 
Because of this, SRS is widely used in studies of metabolic processes \cite{zhang2019spectral}, neuron membrane potentials \cite{tian2016monitoring}, antibiotic responses \cite{schiessl2019phenazine}, nerve degeneration \cite{ tian2016monitoring}, and other biological structures and processes. State-of-the-art SRS microscopes have both reached the shot noise limit \cite{saar2010video,freudiger2008label}, and face photodamage constraints on their illumination intensities \cite{fu2006characterization}, creating a significant roadblock for further advances in capabilities, despite a need for increased performance to enable emerging applications such as high-throughput cancer screening \cite{cheng2021emerging,tan2023profiling}.

One way to overcome the photodamage limit is to introduce a probe field that has quantum correlations, such as squeezed states of light where a non-linear process is used to suppress, or ‘squeeze’, the shot noise \cite{slusher1990quantum,loudon1987squeezed,bowen2023quantum,mauranyapin2022quantum}. This has been demonstrated for SRS imaging \cite{casacio2021quantum,xu2022quantum}. However, in experiments that used a single squeezed beam \cite{casacio2021quantum}, the squeezed light was not bright enough to allow the optimal ratio of Stokes-to-pump power to be achieved. This constrained the SNR and resulted in an imaging frame rate of several minutes. More recently, brighter twin-beam squeezed light fields have been used in SRS imaging, allowing operation significantly closer to the optimal ratio \cite{xu2022quantum,li2022quantum}. However, twin-beam squeezing exploits correlations between two beams, rather than squeezing of a single beam, and this introduces a 3 dB noise penalty \cite{xu2022quantum}. As such, twin-beam experiments have yet to surpass the single-beam shot noise limit of most relevance to SRS imaging.

In this work, we follow the approach of directly generating and detecting a single squeezed field.  We develop a quantum enhanced SRS microscope that is compatible with bright single-beam squeezed light, and use a lock-in amplifier to improve the SNR. We increase the brightness of the probe field compared to previous single-beam SRS imaging experiments \cite{casacio2021quantum}. This allows us to reach the optimal Stokes-to-pump intensity ratio, with light intensities that approach the photodamage threshold of our samples. Combined, optimal Stokes-to-pump ratio and lock-in detection yields an increase of 8.1$\pm$1.7 dB in shot noise limited SNR compared to previous work \cite{casacio2021quantum}. Amplitude squeezing is introduced in the probe field, and it is found that the increased brightness of the probe field degrades the level of squeezing, which we attribute to multi-mode squeezing effects \cite{loudon1987squeezed,kim1994deamplification}. We mitigate this degradation using aperture filtering and report an increase in squeezing level from 0.8 dB to 1.5 dB.

We apply our quantum-enhanced SRS microscope to image living cells, and perform the first demonstration of multispectral quantum-enhanced SRS on living samples. We report an improvement of 1.2 dB improvement over the single-beam shot noise limit when imaging living cells. This quantum enhancement is degraded from the 1.5 dB measured without a sample due, likely, to imperfect suppression of spurious signals that appear at high Stokes powers at the sample. The increase in the SNR enables the imaging speed to be improved while maintaining high image quality with SNR up to 14 dB. We image with a pixel dwell time down to 1 ms, limited by our stage scanning system. This allows 100$\times$100 pixel images to be acquired in 18 s, compared with the 8 minutes required in Ref. \cite{casacio2021quantum}. This faster acquisition time makes the observation of organelle dynamics possible. Through post-analysis of our imaging data, we show that the microscope can resolve SRS signals on living cells with pixel dwell times as low as 3.05 $\unit{\mu s}$, 20\% faster than is possible with coherent illumination. This pixel dwell time is compatible with video-rate (50 Hz) acquisition of 80$\times$80 pixel images.




\section{Fast quantum nonlinear microscope}

\begin{figure}[h]
	\centering
	\includegraphics[width=0.95\columnwidth]{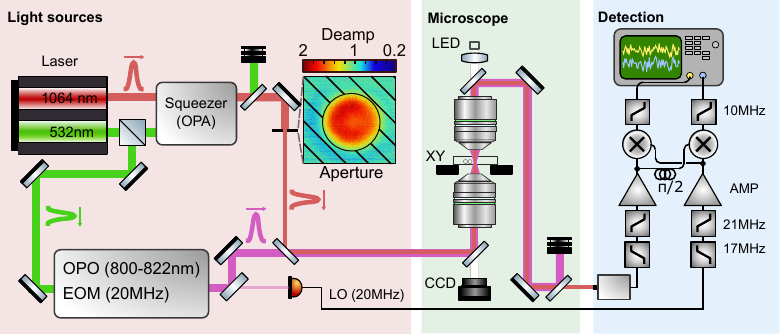}
	\caption{\textbf{Experimental setup}. 
 \textbf{Red Panel:} The SRS pump beam (purple) is generated using an optical parametric oscillator (OPO) and modulated at 20 MHz with an electro-optic modulator (EOM). The Stokes beam (red) undergoes amplitude squeezing in an optical parametric oscillator (OPA) based on second harmonic generation (SHG) within a periodically poled $KTiOPO_4$ crystal, which is pumped with 532 nm light. After passing through the OPA, the squeezed 1064 nm beam is spatially filtered using a tunable aperture to remove the amplified outer ring (as indicated in the inset, where the aperture is represented by hatching). 
 \textbf{Green panel:} The SRS pump and Stokes beams are spatially overlapped and focused to generate the SRS signal within the sample. Bright-field imaging of the sample is achieved using a white LED and a CCD camera. Raster scanning with improved control allows high-speed imaging.
 \textbf{Blue panel:} The 106 4nm light containing the SRS signal is detected by a custom detector, up to 15 mW. A home-built dual quadrature lock-in amplifier digitizes the two quadratures independently and mixes down the SRS DC signals.} 
	\label{fig:fig1} 
\end{figure}

The experimental setup of our quantum-enhanced SRS microscope is depicted in Figure \ref{fig:fig1}. It is similar to, but expands upon, the setup in Ref. \cite{casacio2021quantum}. The light sources were chosen to match those found in high-quality commercial SRS systems. 
We use a pulsed laser with a pulse duration of 6 ps emitting  at 1064 nm and 532 nm with a repetition rate of 80 MHz. The 1064 nm light serves as the Stokes beam, and the 532 nm light is converted into the SRS pump using a tunable optical parametric oscillator (OPO, Levante emerald). The wavelength of the SRS pump can be adjusted between 800 nm and 822 nm, allowing the Raman shift to be scanned across the CH stretch region of the Raman spectra (2800 and 3100 $cm^{-1}$) \cite{camp2015chemically}. The pump beam is modulated at 20 MHz to shift the Raman signal away from low-frequency laser noise and technical noise \cite{freudiger2008label}. This is a standard method in the field of SRS microscopy, and is also used to allow quantum-enhanced particle tracking in optical tweezers \cite{taylor2013biological}.

The Stokes light is amplitude-squeezed using a home-made travelling-wave OPA pumped by a 532 nm beam. The pump and Stokes beams are combined using a dichroic mirror and directed through a pair of high numerical aperture (NA, with NA=1.2) objectives. Both objectives are custom made to allow 92\% transmission at 1064 nm, therefore minimizing the degradation of quantum correlations. The sample is positioned in the focal plane of the input objective, and is placed on a set of piezo-actuated stages 
for raster scanning. The SRS process transfers energy from the pump to the sidebands of the Stokes beam at 20 MHz. After passing through the microscope, the pump beam is filtered out, and the Stokes beam containing both the SRS signal and the quantum correlations is detected on a home-built high-power handling detector resonant at 20 MHz.

Operation of the microscope at the optimal Stokes-to-pump ratio and near-photodamage intensities at the sample requires a high-power handling photodetector as well as generation of bright squeezed light. High-power handling detectors are challenging to design because of saturation effects. The detection of high average intensities results in high photocurrents and puts a constraint on photodiode and detector linearity. Additionally, pulsed light generates transient photocurrents at the laser repetition rate that can saturate the photodetector even when the equivalent mean intensity is below the saturation level of the detector \cite{guay2021balanced}. Generating bright squeezed light is also a challenge because an increase in brightness of the squeezed field has been associated with a degradation in the level of squeezing that can be measured \cite{PhysRevLett.59.2566,PhysRevLett.70.1244,PhysRevA.45.458}, attributed to multimode effects.
In the following three sections we present the solutions that we implemented to overcome these challenges.

\FloatBarrier



\subsection*{Detection system}

Limitations in detector power-handling due to saturation effects have been one factor that has prevented previous quantum-enhanced SRS demonstrations from operating at the optimal Stokes-to-pump power. In reference \cite{casacio2021quantum}, the Stokes power was limited to 3 mW due to nonlinearity in the detector's response that occurred at higher powers. To operate close to the photodamage threshold on biological samples, where quantum enhancement is relevant, a Stokes-to-pump intensity ratio of 0.1 was used, significantly below the optimum ratio of 0.5. Similarly, in reference \cite{xu2022quantum}, the SRS signal is detected on the pump beam, leading to an optimum Stokes-to-pump intensity ratio of 2, while the power-handling of the detector constrained the Stokes-pump ratio to approximately 0.85.

Here, we develop a high quantum efficiency, high-power handling detector that is linear up to a power of 15 mW of pulsed light. We use a 500 $\mu m$ diameter high-efficiency photodiode (IG17X500S4i, Laser Components) and  resonant feedback to address the saturation challenge originating from the high mean and peak light intensity. The photodiode was chosen for its linearity at high intensities compared to Ref. \cite{casacio2021quantum}, along with its low capacitance and dark noise. The resonant feedback includes a bandpass filter centered around the SRS modulation frequency. This maximises the gain at the SRS modulation frequency while suppressing the photodiode response at DC and at the laser repetition rate that would otherwise cause detector saturation. The quantum efficiency of the photodiode is measured to be 82\%, limiting the photodetector quantum efficiency. The efficiency was determined by measuring the photocurrent $i$ produced by 15 mW of Stokes light and comparing the flux of electrons associated with $i$ to the flux of photons arriving at the detector.

\begin{figure}[h]
	\centering
	\includegraphics[width=0.5\columnwidth]{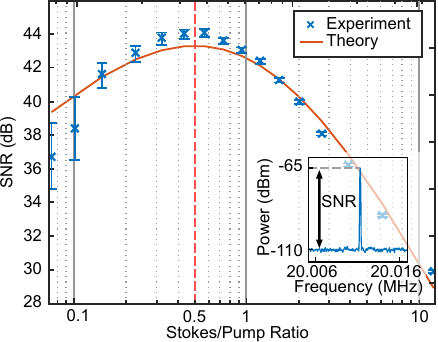}
	\caption{\textbf{Optimization of the signal-to-noise ratio (SNR).}
 Evolution of the SNR of the SRS process as a function of the Stokes/pump power ratio for a fixed total power at the sample. The blue data points represent the experimental measurements with electronic noise subtraction, and the orange line is the theoretical prediction. The vertical red line highlights the optimal Stokes-to-pump ratio, associated with the highest SNR. Inset: typical SRS spectra (corresponding here to a Stokes-to-pump ratio of 0.56), the SNR is obtained by subtracting the noise floor to the peak value of the SRS.
 } 
	\label{fig:SNR_imp} 
\end{figure}

 We characterize the SNR dependence to the Stokes-to-pump intensity ratio. To do so, we perform the SRS measurement on silica beads at a Raman shift of 3055 cm$^{-1}$. Then, we monitor the SNR of the SRS power spectrum as we vary the Stokes-to-pump intensity ratio. The inset in Figure \ref{fig:SNR_imp} shows a typical SRS power spectrum corresponding to a Stokes-to-pump ratio of 0.56. The peak corresponds to the SRS signal, and the SNR is calculated by subtracting the peak value to the noise floor. 
The Stokes-to-pump intensity ratio is varied between 0.07 and 10, with the total intensity at the sample kept constant at 75 $\unit{W \mu m^ {-2}}$, beneath the photodamage threshold around 80 $\unit{W \mu m^ {-2}}$ \cite{casacio2021quantum}. The variation of the SNR with the Stoke-to-pump intensity ratio is presented in Figure \ref{fig:SNR_imp} in blue, with the orange trace showing the theoretical prediction \cite{cheng2016coherent}. There is good agreement between theory and experiment,  confirming that the optimal Stokes-pump ratio is around 0.5. In addition, a 5.5 $\pm$ 1.7 dB improvement in SNR is observed between an optimum 0.5 intensity ratio compared to the 0.1 ratio used in Ref \cite{casacio2021quantum}. The experimental uncertainty in the data points stems from electronic pick-up fluctuations, which are particularly noticeable at low Stokes power.

The high-power handling detector allows the microscope to operate at the optimal Stokes-to-pump ratio, with up to 17.8 and 35.6 mW of Stokes and pump power at the sample, respectively. This total power matches the power levels ($>$30 mW) used in state-of-the-art SRS microscopes \cite{freudiger2008label,shou2021super}, and corresponds to an intensity at the sample of approximately 230 $\unit{W \mu m^ {-2}}$. 

To further improve the SNR, we introduce a lock-in amplifier to our detection scheme. Lock-in amplifiers have been used by the radio frequency community to extract weak signals from noisy backgrounds \cite{meade1982advances,dixon1989broadband}. They have become a standard approach in SRS microscopy \cite{freudiger2008label, nandakumar2009vibrational, zhang2011highly,xu2022quantum} and have been applied to other precision light microscopy techniques \cite{taylor2013biological,taylor2013optical,mauranyapin2017evanescent}.
A lock-in amplifier uses a local oscillator (LO) that matches the frequency and phase of the signal to only measure the quadrature of the photocurrent that contains the signal. This phase referencing leads to a 3 dB improvement in SNR.

The design of our lock-in amplifier detection is presented in Figure \ref{fig:fig1} (“Detection” panel). Typically, the LO is generated electronically; however, here, the LO needs to precisely match the modulation frequency of the SRS signal that was found to vary slowly with time. We implement precise frequency-matching by optically detecting the pump beam at the back of a polished mirror after the EOM, as shown in Figure \ref{fig:fig1}. The LO and signal photocurrents are bandpass-filtered to retain the 20 MHz component and amplified before being mixed down to separate the two quadratures. The LO phase is optimised so that the SRS signal is only present on one quadrature, and is then digitized on an oscilloscope. 
We report a 2.6 dB improvement in the SNR by introducing the lock-in amplifier compared to the direct detection scheme. This is in good agreement with the theoretical 3 dB, and we attribute the the deviation to electronic noise. 
Combined, operating at the optimal Stokes-pump ratio and employing lock-in detection allowed an 8.1 $\pm$ 1.7 dB improvement in shot noise-limited SNR compared to the best previous single-beam squeezed Raman microscope \cite{casacio2021quantum}.

\subsection*{Addressing electronic pick-up and spurious signals}
 With the detection improvements, we found that the microscope became significantly more sensitive to deleterious effects that were previously buried under the noise floor. First, the level of electronic pick-up was increased. The pick-up was addressed by carefully isolating the detection components and packaging the detector and its power supply within a Faraday cage. Second, the increase in Stokes power at the sample led to spurious signals competing with the SRS. These signals have been reported in other high-power SRS microscopes, and can originate from nonlinear effects such as Kerr modulation and two-color two-photon absorption \cite{genchi2023background}. We found that they could be suppressed over the imaging time period by adjusting the pump beam polarization.


\FloatBarrier

\subsection*{Addressing spatial mode effects in bright squeezed light generation}

To perform quantum-enhanced SRS, we pump the OPA in the Stokes arm of the experiment to deamplify the seed and therefore produce amplitude squeezing. As we increased the OPA seed power above those used in Ref. \cite{casacio2021quantum} to achieve brighter squeezing, we observed a degradation in the level of squeezing, similar to that reported in Refs. \cite{PhysRevLett.59.2566,PhysRevLett.70.1244,PhysRevA.45.458}. This degradation has been attributed to \textit{gain-induced diffraction} (GID) \cite{kim1994deamplification}. In GID, the radial intensity profile of the OPA pump induces a phase shift in the OPA process, leading to less deamplification or even local amplification of the OPA seed. Another cause of squeezing degradation is spatial mode mismatch, where the Gouy phase shift introduces a phase shift between the on- and off-axis components of the seed and pump fields, leading to reduced deamplification or amplification of off-axis components \cite{lassen2007tools}.

We note that direct-detection experiments such as ours are more sensitive to local amplification in the seed beam than experiments that use homodyne detection. On the one hand, regions of the seed that are amplified have both their intensity and noise increased, resulting in a double amplification of the detected noise (since the detected noise is proportional to the product of the two). On the other hand, regions of the seed that are deamplified and contain squeezing have their intensity decreased, reducing the contribution of the squeezed part of the field to the detected signal.

\begin{figure}[h]
	\centering
	\includegraphics[width=0.95\columnwidth]{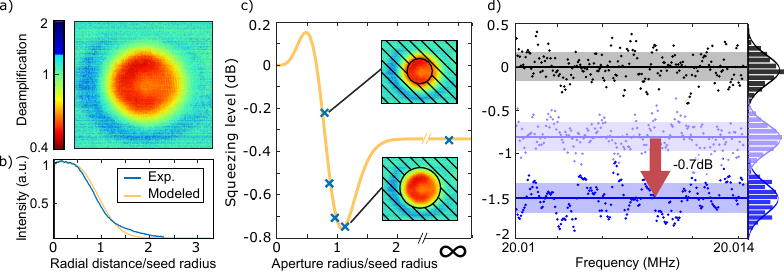}
	\caption{\textbf{Squeezing improvement via spatial filtering.} 
 \textbf{a)} Ratio of intensity between the signal beam with and without pump. The red region (ratio $>$1) corresponds to deamplification of the beam and squeezing, while the blue region (ratio $<$ 1) corresponds to amplification and anti-squeezing.
 \textbf{b)} Normalised radial intensity profile of the seed beam in presence of the pump, experimental data (blue trace) and fit with our model (yellow trace).
 \textbf{c)} Blue data points: experimental level of squeezing as a function of the radius of the aperture filter. Yellow trace: modeled squeezing level. Insets show the filtered area of the beam for two aperture radii, and the corresponding squeezing level.
 \textbf{d)} Comparison of the squeezing level with no aperture (light blue) and with the optimised aperture radius (blue) relative to the shot noise level (black). A 0.7 dB improvement is reached using the aperture filtering with a radius of 1.1 relative to the Stokes beam radius.}
	\label{fig:SQ_imp} 
\end{figure}

To investigate whether the deamplification of the Stokes field is uniform across its transverse profile, we measure the intensity profile of the beam with and without squeezing on a CCD. The ratio of the two intensities gives the deamplification factor. As shown in Figure \ref{fig:SQ_imp}(a), a deamplification factor above one is measured at the center of the beam, indicating that the OPA in this region works as intended. However on the outer part of the beam we observe local amplification, explaining the degradation in the squeezing level.


Spatial filtering has been proposed in theoretical studies of amplitude-squeezing \cite{koprulu2001analysis} to mitigate GID and Gouy phase shift effects. Here, we implement an experimental demonstration. An aperture is placed on the Stokes beam after the OPA to remove the amplified fraction of the beam, as shown in Figure \ref{fig:fig1} in the 'Light sources' panel. 
We measure the squeezing level as the aperture radius is varied and plot the result in Figure \ref{fig:SQ_imp}(c) (blue data points). We find that for aperture radii between 1.8 to 1 times the seed beam radius, the squeezing level is improved. As the aperture radius is further reduced, the squeezing level drops and eventually becomes worse than the apertureless process. The two insets in Figure \ref{fig:SQ_imp}(c) illustrate cases where the aperture enhances the squeezing level (for a radius of 1.1) and where the aperture introduces excessive losses (for a radius of 0.8). The optimal aperture radius is found to be approximately 1.1 times the seed beam radius and allows for 1.5 dB of squeezing to be observed, increased from 0.8 dB with no spatial filtering. The improvement is squeezing level is illustrated in Figure \ref{fig:SQ_imp}(d) where the noise spectrum is plotted for shot noise (black), squeezing with no aperture (light blue) and squeezing with optimal aperture radius (bright blue).

To obtain a qualitative understanding of the effect of spatial filtering, we develop a simple phenomenological model of the OPA process that accounts for local beam amplification. We decompose the output field of the OPA into a set of Laguerre Gaussian (LG) modes, since the field is cylindrically symmetric. We restrict the LG decomposition to the two first modes $\{u_{00}(r), u_{01}(r)\}$ of width $w_0$ (normalized by $w$, the width of the OPA seed), similar to previous theoretical work \cite{koprulu1999analysis,kawase2008observing}. We expect the dominant multimode effects to be accounted for by two modes, and therefore neglect the higher order modes. In addition, we assume that the two modes are in-phase. This is consistent with the shape of the pumped output field, which exhibits a flat top (see Figure \ref{fig:SQ_imp}(b), blue trace).
The OPA process is then modeled as applying deamplification factors $A_{00}$ and $A_{01}$ on the first and second LG components, respectively. Finally the aperture is introduced as a low-pass radial filter, and the level of squeezing in the filtered beam is calculated, taking into account the losses introduced by the aperture as well as the detection efficiency of the setup $\eta$, measured experimentally to be 0.55.  

We fit our model to the experimental intensity profile of the OPA output in presence of the pump using $w_0$, $A_{00}$ and $A_{01}$ as fitting parameters. Good agreement between experimental and modeled intensity profiles is observed for $[w_0, A_{00}, A_{01}]=[0.855,0.6,3]$ as shown in Figure \ref{fig:SQ_imp}(b). Minor deviations at high radial distances are noted, and are likely due to contributions from higher order LG modes. The squeezing level is then calculated as a function of the aperture radius with no additional fitting parameters and plotted in Figure \ref{fig:SQ_imp}(c) (yellow trace).

Although the model is simple, good agreement is observed between the experimental data and the model prediction in Figure \ref{fig:SQ_imp}(c), indicating that it captures the dominant physics.  The decomposition of the output OPA field indicates that only 2.4\% of the beam intensity is amplified, however the amplification factor $A_{01}$ of 3 is significantly higher than the deamplification factor ($A_{00}$=0.6) and explains the strong squeezing improvement enabled by the aperture filtering. Interestingly, for high levels of aperture filtering the model predicts an increased variance of the beam compared to the shot-noise level (0 dB of squeezing). This corresponds to a regime where the aperture radius removes most of the deamplified component and the noise originating from the central part of the amplified component of the beam becomes dominant. This regime was not observed experimentally because the required aperture radius was beneath the minimum radius available with the aperture.

\FloatBarrier

\section{Quantum-enhanced imaging of living cells}

\begin{figure}[h]
	\centering
	\includegraphics[width=0.65\columnwidth]{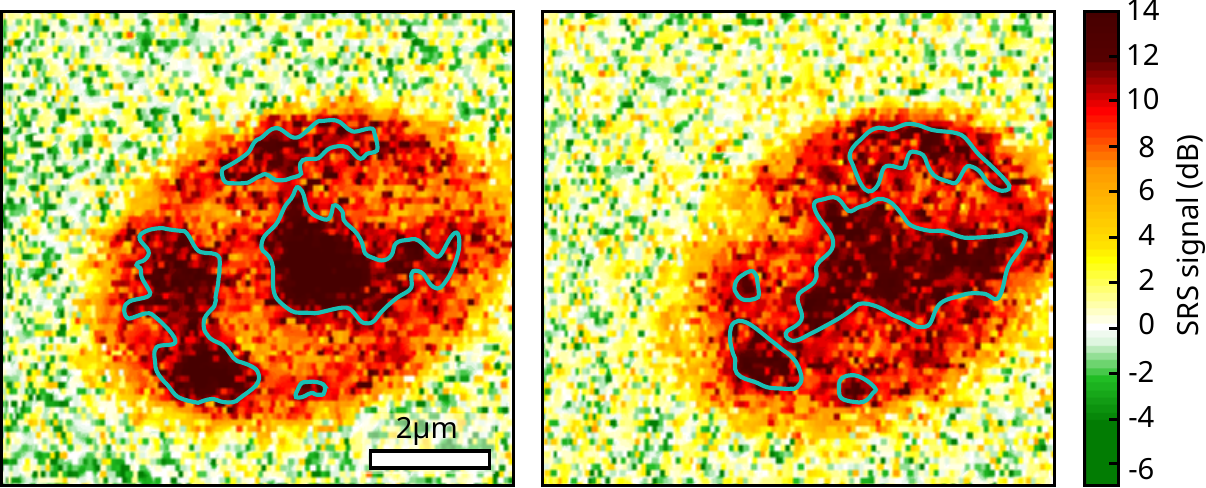}
	\caption{\textbf{Quantum-Enhanced Imaging of Organelle Motion}: Quantum-enhanced SRS images of a yeast cell at a Raman shift of 2850 $\unit{cm^{-1}}$ (predominantly originating from lipids). The respective quantum enhancement levels are 1.2 dB for the left image and 0.7 dB for the right image. The blue contour indicates the regions of the cell where the SRS signal exceeds 10.5 dB and highlights cell structure and organelles. 
 }
	\label{fig:orga_dyn} 
\end{figure}

We apply our microscope to perform fast quantum-enhanced imaging of living cells. 
In Figure \ref{fig:orga_dyn}, we present two sub-shot noise images of living yeast cells (\textit{Saccharomyces cerevisiae}, acquired at a Raman shift of 2850 $\unit{cm^{-1}}$ that is associated with lipids \cite{lu2015label}. The cells were rehydrated in distilled water mixed with sugar and NaCl salt (0.1\% by weight) for two hours, then placed between two coverslips. The images are acquired by raster scanning over a 10$\times$10 $\mu m$ area, with a pixel dwell time of 1 ms and pixel size of 100 nm. The full image takes 18 seconds to be acquired, maintain an SNR of up to 14 dB, and display a clear definition of the cell boundary as well as inner features. The image is of high quality compared to other coherent Raman images of yeast cells \cite{svedberg2010nonlinear, okuno2010quantitative, furuta2017intracellular,kochan2018single,lima2021spectral} despite using a much shorter pixel dwell time (1 ms compared with 50 ms). We note that while our images are acquired significantly faster than typical yeast cells images, the scanning speed used is limited by the mechanical response of our piezo system, and remains lower than other demonstration of quantum-enhanced SRS imaging of polymers \cite{xu2022quantum} or state-of-the-art SRS microscopes \cite{freudiger2008label} that can achieve dwell times down to hundreds of $\mu s$ per pixel on specimens with high Raman contrast.

The quantum enhancement is found to be 1.20 dB and 0.72 dB for the left and right image in Figure \ref{fig:orga_dyn}, respectively, and was determined by comparing the noise in a region of the background on the squeezed images to a reference shot noise image of the same region with no SRS pump and the squeezer off. Compared to Ref \cite{casacio2021quantum}, the level of quantum enhancement varied between the images due to drift in the suppression of the spurious background discussed previously. The reduction in squeezing is more than compensated for by the increase in SNR enabled by working at the optimal Stokes-pump ratio and by employing a lock-in detection scheme. Importantly, since direct detection of a bright squeezed beam does not require a difference-squeezing measurement between two detectors, it avoids the 3 dB noise penalty present in other recent experiments \cite{xu2022quantum,li2022quantum}.

Figure \ref{fig:orga_dyn} reveals information about the cell structure. The Raman signal from the cytosol is visible across the whole cell, while several bright organelles can be observed. The outer organelles likely correspond to lipid granules that form in yeast cells \cite{kochan2018single}, while the main central feature is compatible with the nucleus. Lipid granules are expected to be approximately spherical. Here, the distorted shapes of the bright features can originate from droplets being out of focus.

The imaging speed opens up the possibility to observe medium and short-term biodynamics.  
We apply this capability in the images acquired in Figure \ref{fig:orga_dyn}. The two images were acquired three minutes apart, which was the delay required to observe meaningful changes in organelle distribution of this particular cell. To highlight the position of the organelles, we introduce blue contours around the areas where the SRS signal is above 10.5 dB. As can be seen, the bottom left feature in the ealier image (F)igure \ref{fig:orga_dyn}, left) has split in the second image (Figure \ref{fig:orga_dyn}, right) and appears to overlap with the central feature, while the top feature has moved to the right by about 2 $\unit{\mu m}$.

\FloatBarrier

\begin{figure}[h]
    \centering
    \includegraphics[width=0.7\columnwidth]{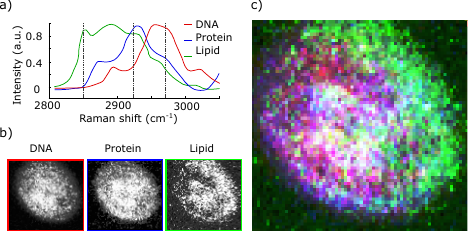}
    \caption{\textbf{Multispectral Imaging}: \textbf{a)}: SRS intensity of pure DNA, protein, and lipid as a function of Raman shift, extracted from \cite{lu2015label}.
    \textbf{b)}: Quantum-enhanced SRS images of a yeast cell at the Raman shifts 2967, 2926, and 2850 $\unit{cm^{-1}}$, predominantly originating from DNA, proteins, and lipids, respectively. The corresponding levels of quantum enhancement for the three images are 0.91 dB, 1.12 dB, and 0.92 dB. \textbf{c)}: Overlaid RGB image of the yeast cell obtained by combining the three individual shift images. The red, blue and green channels correspond to DNA, proteins and lipids, respectively.}
    \label{fig:multi_sp}
\end{figure}

In addition to observing organelle dynamics, the speed of our setup allows us to conduct quantum-enhanced multispectral scans across a range of Raman shifts, reported here for the first time on a biological sample. Within the CH/OH stretch region, we identify three specific Raman shifts: 2967, 2926, and 2850 $\unit{cm^{-1}}$, predominantly originating from DNA, proteins, and lipids, respectively \cite{lu2015label,camp2015chemically}. These contributions of DNA, proteins, and lipids are plotted against the Raman shifts in Figure \ref{fig:multi_sp}(a).

Quantum-enhanced imaging is performed at each shift on a yeast cell, and the resulting images are depicted in Figure \ref{fig:multi_sp}(b). Each composite image is renormalized between 0 (black) and 1 (white). The respective quantum enhancements are 0.91 dB, 1.12 dB, and 0.92 dB. Some distinctive features can be identified in the individual shift images: the DNA image displays a strong region between 11 and 13 dB of SNR slightly below the center of the cell (likely the nucleus), while the lower signal present in the rest of the cell is consistent with RNA molecules, which exhibit a similar Raman signature to DNA \cite{prescott1984characterization}.
 The lipid image shows a peak SNR of 9.5 dB around the same position as the DNA image, possibly originating from the lipid present in the nuclear membrane and in the nucleus. Another high SNR region of the SRS is located near the right and top of the cell, potentially from out-of-focus lipid granules. 
Last, the protein image exhibits a robust signal covering most of the cell, and is on average higher than the DNA image SNR by 3 to 5 dB and peaking at 18 dB of SNR. Higher SNR is expected because protein concentrations are higher than DNA and lipids across the majority of the cell \cite{siwiak2010comprehensive}. 

To visualize these shifts concurrently, we overlay them on the RGB channels of a common image in Figure \ref{fig:multi_sp}(c) where the red, blue and green channels correspond to DNA, proteins and lipids, respectively. The correlation of all three shifts at the center of the cell associated with the nucleus appears clearly, as well as the distinct spatial separation between the top-right region where the shift associated with lipids dominate and the DNA and protein signals are weaker.

Despite the features we have analyzed, some elements do not match the expected distribution of chemicals within the yeast cells. Notably, the DNA and protein images are closely correlated over the cell, whereas the concentration of protein is expected to be lower than that of DNA in the region associated with the nucleus. This difficulty to separate each individual component and perform linear decomposition stems from the relatively small size of yeast cells (5 to 7 $\mu m$ in diameter combined with flattening as they weakly bind to the cover slip) while HeLa, for example, used in previous work \cite{lu2015label} have diameters between 20 and 30 $\mu m$. This limited height combined with the axial resolution of the focal beams, estimated at 800 nm from the product of the Rayleigh length of the Stokes and pump beams \cite{cheng2021emerging}, means that several organelles are likely to be averaged in the signal from a single pixel. For instance in the DNA image, the SRS signal from the central area of the cell, where the nucleus is inferred to be located, also likely includes a fraction of the cytoplasm, causing a fraction of protein to be mixed in the SRS response.


\FloatBarrier

\section{Minimum pixel dwell time compatible with video rate imaging}

\begin{figure}[h]
	\centering
	\includegraphics[width=0.65\columnwidth]{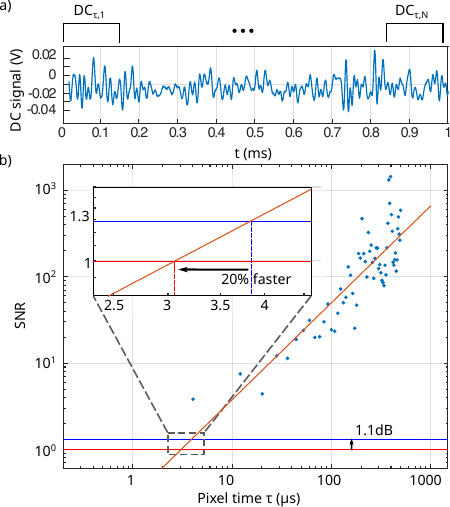}
	\caption{\textbf{Estimation of minimum pixel dwell time.} 
 \textbf{a)}, Example of the DC trace of a pixel with SRS signal when the squeezer is on (extracted from a quantum enhanced image at the 2926$\unit{cm^{-1}}$ Raman shift). The total acquisition time at the pixel is 1 ms, the brackets at the top illustrate how shorted dwell times are simulated by dividing the trace into shorter subsets of time, $\tau$. 
 \textbf{b)}, Plot of the simulated SNR versus pixel dwell time (blue dots) accompanied by a linear fit (orange). A horizontal red line is drawn at SNR=1 and its intersection with the fit indicates the  minimum pixel dwell time (3.05 $\unit{\mu s}$). Additionally, the blue line, placed 1.1dB above (SNR=1.29), corresponds to the shot noise-limited minimum pixel dwell time (3.85 $\unit{\mu s}$).
}
	\label{fig:fig4} 
\end{figure}

The images shown in Figure \ref{fig:orga_dyn} and \ref{fig:multi_sp} display high SNR, with the protein image from Figure \ref{fig:multi_sp} peaking at 1000, which means that faster imaging should be possible at the expense of some SNR. The scanning stages used in this work limit raster scanning speed to 1 ms per pixel. Therefore, we simulate shorter pixel dwell times to extract the minimum pixel dwell time corresponding to a SNR of 1.
We accomplished this by selecting a quantum-enhanced SRS time-trace obtained from living yeast cells, associated to a pixel with strong SRS at the Raman shift $2926\unit{cm^{-1}}$ (protein shift). 
From the 1 ms-long trace, we simulated shorter pixel dwell times $\tau$ by dividing the SRS timetrace into $N$ $\tau$-long segments, illustrated in Figure \ref{fig:fig4}(a). The $N$ segments are used to estimate the SNR associated with a pixel dwell time $\tau$:
$SNR_{\tau}=\bar{DC}_{\tau,1}^2/std(\bar{DC}_{\tau,1},...,\bar{DC}_{\tau,N})^2$, where $\bar{DC}_{\tau,i}$ is the mean value of the SRS signal timetrace of the $i$-th segment and $std$ is the standard deviation. 

In Figure \ref{fig:fig4}(b) we plot the simulated SNR as a function of pixel dwell time $\tau$ (depicted by blue points). We fit a line of slope 1, in orange, as the SNR scales linearly with pixel dwell time. The minimum pixel dwell time is estimated to be 3.05 $\mu$s by intersecting the line fit with the horizontal line at SNR =1. This minimum time includes a measured 1.1 dB of squeezing. Without the 1.1dB of quantum enhancement, the shot-noise limited minimum pixel dwell time is found to be 3.9$\mu$s. Therefore, the quantum enhancement provides a 20\% improvement in minimum pixel dwell time. 
Importantly, these times are compatible with video-rate imaging, enabling the capture of 80$\times$80 pixel image at a standard rate of 50 Hz, or 71$\times$71 images at the same rate without quantum enhancement. In the future, using a faster raster scanning method, such as a set of galvanometer mirrors
will allow quantum-enhanced video-rate imaging of biological samples.

\FloatBarrier

\section{Conclusion}

In this work, we have developed a quantum-enhanced SRS microscope using bright, single-beam squeezed light that operates at the optimal SRS intensity ratio. The increased illumination power caused a degradation in the squeezing level due to multimode effects and local amplification of the field. We mitigated these effects by spatially filtering the Stokes beam, which improved the squeezing level from 0.8 dB to 1.5 dB. Between classical and quantum optimisation, an increase in SNR of 8.8$\pm$1.7 dB is measured compared to the best previous single-beam quantum-enhanced SRS microscope \cite{casacio2021quantum}. These improvements allow the living cells to be imaged with pixel dwell time down to 1 ms while maintaining a SNR up to 14 dB and an observed quantum enhancement of 1.2 dB compared to the single-beam shot noise limit. The microscope enables the observation of organelle dynamics and multispectral quantum-enhanced SRS on living cells for the first time. Additionally, we showed that with faster scanning, the SRS signal can be resolved on 80$\times$80 pixel images at a 50Hz frame rate, with quantum correlations yielding a 20\% increase in speed.



\section{Acknowledgements}
This research was supported by the Air Force Office of Scientific Research under award numbers  FA9550-20-1-0391 and FA9550-22-1-0047, the Australian Research Council Centre of Excellence for Engineered Quantum Systems (EQUS, grant number CE170100009) and the Australian Research Council Centre of Excellence in Quantum Biotechnology (QUBIC,grant number CE230100021).

\bibliographystyle{tfnlm}
\bibliography{interactnlmsample}

\end{document}